# User-level Weibo Recommendation incorporating Social Influence based on Semi-Supervised Algorithm

Daifeng Li[1], Zhipeng Luo[2], Golden Guo-zheng Sun[3], Jie Tang[1], Jingwei Zhang[4]

[1]Dept. of Computer Science and Technology, Tsinghua University, Beijing, China
[2]Beijing University of Aeronautics and Astronautics, Beijing, China
[3]Tencent Company, Beijing, China
[4]Department of Electronic Engineering, Tsinghua University, Beijing, China

ldf3824@yahoo.com.cn, patrick.luo2009@gmail.com, gordon.gzsun@gmail.com, jery.tang@gmail.com, iceboal@gmail.com

**Abstract:** Tencent Weibo, as one of the most popular micro-blogging services in China, has attracted millions of users, producing 30-60 millions of weibo (similar as tweet in Twitter) daily. With the overload problem of user generate content, Tencent users find it is more and more hard to browse and find valuable information at the first time. In this paper, we propose a Factor Graph based weibo recommendation algorithm TSI-WR (Topic-Level Social Influence based Weibo Recommendation), which could help Tencent users to find most suitable information. The main innovation is that we consider both direct and indirect social influence from topic level based on social balance theory. The main advantages of adopting this strategy are that it could first build a more accurate description of latent relationship between two users with weak connections, which could help to solve the data sparsity problem; second provide a more accurate recommendation for a certain user from a wider range. Other meaningful contextual information is also combined into our model, which include: Users□ profile, Users□ influence, Content of weibos, Topic information of weibos and etc. We also design a semi-supervised algorithm to further reduce the influence of data sparisty. The experiments show that all the selected variables are important and the proposed model outperforms several baseline methods.

## Categories and Subject Descriptors
H.3.3 [Information Systems]: Information Search and Retrieval: Information Filtering

## General Terms
Algorithms, Experimentation

## Keywords
Tencent, Weibo Recommendation, Factor Graph, Topic Model, Social Influence

## 1. INTRODUCTION
Tencent is the most popular microblogging service in China, it is an important platform combining both social media and social network, which contains 300 million users as of 2011. It allows users to share information with their followers or the public by posting messages of up to 140 Chinese characters, which are called weibos. Average 30-60 million weibos are generated per day. Users can access all weibos generated by a specific person, and forward weibos to share it with his/her friends. The forward behaviors could accelerate the spread of information in the social network than traditional social media. Many Tencent users take Tencent as a personalized media center, which could provide them the newest information about political events, economics, celebrities, their friends at the first time, and Tencent Weibo has become one of the most important information source in their daily lives.

However, as a result of the rapidly increasing number of weibos, most Tencent users encounter a serious problem of information overload; according to our statistical analysis, Tencent users follow 64 people on average, who will generate hundreds or even thousands of weibos each day. This is inconvenient for users to browse and find useful information, especially for those active users, who often have more followees than others, they often spend more time to check all weibos to find useful information. Chen et al [1] provide more detailed descriptions of that problem. Thus, recommending useful weibos to a user is an important challenge and the focus of this paper. Intuitively, a weibo is useful to a user, if the user is interested in or willing to read the weibo. Whether a user is interested in a tweet is determined by many factors, such as the quality of the weibo, the influential degree of the authors, etc. Personal interest is also an important factor to decide whether a tweet is personally useful. Chen et al[1] considered topic factor, social relations, users' interesting preferences and etc to generate a collaborative ranking framework. Yan et al[2] proposed a graph-theoretic model for tweet recommendation, they generate three networks to connect users, items together to observe the influence of users' preferences, popularity, diversity and influence of tweets and etc. Similar with their researches, our work also consider those important factors, such as Topic information, users' profile from their historical record, the influence of a user, the abstracted key content from a weibo and etc. The most important difference is that our research also incorporates social influence into consideration; the direct influence is studied by the daily communication between two users, while the indirect influence is learned by applying social balance theory [3][4][5], and we also consider to constraint the influential relationship under different topics. The main contributions of the paper are concluded as follows:

- Our model incorporates explicit Tencent features such as the influential degree of users, the topic information, main content of weibos, social relations and topic information into a unified framework, which could further help to improve the recommendation results.
- For direct social influence, we learned the influential relationship between two users by studying their historical communication records.
- For indirect social influence, we learned the influential relationship by applying social balance theory.
- We verified our proposed model on a large scaled of Tencent Data, which could help us better understand users' behaviors in Tencent weibo.

## 2. RELATED WORK

### 2.1 Twitter Recommendation

As Twitter has become a popular social medium and had great impact, plenty of researches focus on analyzing the personal interest of users and building recommendation algorithms. Michelson et al [6] detect the entities of each tweet, and discover the topics of interests for Twitter users. Ramage et al [7] applied Labeled topic models to analyze the content of each tweets. Yang et al [8] established a joint friendship-interest propagation model to make link prediction and tweet recommendation in a unified framework. Chen, et al[1] proposed a collaborative personalized tweet recommendation algorithm, which integrate Twitter factors into a unified framework. Yan, et al [2] made Tweet Recommendation by combining three graphs together. Those researches did not take deep insight into how social influence is generated according to users' historical record and how the influence will determine the results of tweet recommendation. In this paper, we combined both global Tencent features and topic level social influence into a unified framework, which is proved to gain a better performance than traditional method.

**2.2 Social Influence**

One main purpose of social influence analysis is to detect and evaluate the existence of social influence [9]. In addition, Kempe et al. [10] constructed a NP-Hard problem to solve influence maximization in social network. Tang et al. [11] learned social influence on different topics and proposed Topical Affinity Propagation (TAP) to model topic-level social influence. Liu et. al [12] designed a LDA based Social Influence model to detect influential relationship among individuals. Crandall et. al [13] developed techniques for identifying and modeling the interactions between social influence and selection by using data from online communities.

**2.3 Factor Graph**

Factor Graph is a probability based graph model, which is generated by Bayesian network or Markov random fields [14]. The Factor Graph is operated passing "message" along the edges of the graph. Factor Graphs are mainly used to model complex real-world systems and to derive practical message passing algorithms for the associated detection and estimation problems [15]. In recently years, Factor Graph is also widely used in different kinds of social networks, such as Twitter [16,17], Academic Search [18], PatentMiner [19]. In this paper, we use Factor Graph to analyze the communication networks, which are generated by Tencent users, we also make extension for the traditional Factor Graph to realize social influence analysis, which abstract a influential edge into a point and incorporate "Social Balance Theory" into it. The improved methods could capture the influential relationship easier and more efficient. The experiment results also show that incorporating social influence could significantly improve the performance.

## 3. PROBLEM DEFINITION

In this section, we first give several necessary definitions and then present a formal definition of the problem.

A static social network can be represented as $G=(V, E, I)$, where $V$ is the set of $|V|=N$ users and $E \subset V \times V$ is the set of directed links between users. In this paper, we only consider the "Forward" relationship as the links among users, which is based on the pre-assumption: "The user has a high probability to be interested in a weibo, if he/she forward this one". $I$ is the set of original items/weibos. Given this, we can define the user's action as follows.

*Definition 1.* **Attributes of weibos/items:** The main attributes of a weibo/item $X$ can be described as $X=\{Username, KW\{K_1:W_1; K_2:W_2; K_3:W_3\}, TP, Time\}$, where $KW$ is the set of most important key words list of weibo/item $X$, $K_i:W_i$ represents the *ith* key word $K_i$ and its weight $W_i$ in $X$. The extraction of key words and the calculation of weights can be applied by using FudanNLP[1]. *Username* is the author ID of weibo/item $X$.

*Definition 2.* **Attributes of users:** When recommending user $A$'s weibo to user $B$, user $A$'s attributes towards user $B$ are also important for recommendation prediction. In our research, the main attributes are mainly about: $GN$(The number of total replies and comments and mentions and forwards of $A$'s followers); $RN$(The number of weibos-replies between $A$ and $B$); $CN$(The number of weibos-comments between $A$ and $B$); $FN$(The number of weibo-forwards between $A$ and $B$); $MN$(The number of weibo-mentions between $A$ and $B$);

*Definition 3.* **Direct Influence between users:** The topic level influence of $A$ towards $B$ $D\_IN_{A->B}^k$ can be defined as how $B$ will be influenced by $A$ on topic $k$. The range of $D\_IN_{A->B}^k$ is from –1 to 1, where $D\_IN_{A->B}^k<0$ means $A$ has a negative influence on $B$, $D\_IN_{A->B}^k>0$ means $A$ has a positive influence on $B$. Negative means that $B$ has a high probability to dislike $A$'s weibo on topic $k$; Positive means that $B$ has a high probability to like $A$'s weibo on topic $k$. Direct Influence means that $IN_{A->B}^k$ is learned through the communication record of $A$ and $B$.

*Definition 4.* **Indirect Influence between users:** Indirect Influence can be defined by applying Social Balance theory[18], according to Social Balance theory, we define indirect influence as if user $B$ likes $A$'s weibos related with topic $k$, $A$ likes $C$'s weibos related with topic $k$, then $B$ has a high probability to like $C$'s weibos related with topic $k$.

Deninition 5. **Weibos Recommendation:** for user $B$, we define the format of a recommended weibo as $W=\{Z, username\{GN,RN,CN,FN,MN\}, KW\{K_1:W_1; K_2:W_2; K_3:W_3\}, TP, Time\}$.

## 4. PROPOSED MODEL

Based on the above intuitions, we propose a TSI_WR model to learn Tencent users' behaviors and make recommendation. Assume we have $U$ users and $M$ weibos The objective function is defined as below:

$$P(Y|G) = \prod_i \prod_j f(y_{ij}, Z_i, X_j(U = user\_attribute))g(y_{ij}, Z_i, X_j(TP))h(y_{ij}, Z_i, X_j(KW))$$

(1)

where $Y=\{y_{11}, y_{12},...y_{UM}\}$ represents the results of recommendation, $y_{ij}$ represents user $i$ likes the weibo $j$. $Z_i$ represents the latent attributes matrixes of user $i$. $f$, $g$ and $h$ are functions of conditional probability distribution of $P(y_{ij}|X_j(user\_attribute))$,

---
[1] http://code.google.com/p/fudannlp/downloads/list

$P(y_{ij} | X_j(TP))$, $P(y_{ij} | X_j(KW))$. In order to build the model as simple as possible, we abstract each "Recommending Behavior" as a node, and design Factor Model based on those abstraction. According to our statistical analysis based on three months Tencent data, we found the distribution of members forwarding a user's weibos satisfies a power law distribution, which could be described as in figure 1; it illustrates the characters of tip peak and fat tail, which means that a few number of users will frequently forward target users' weibo, while the majority have a low activity to do that.

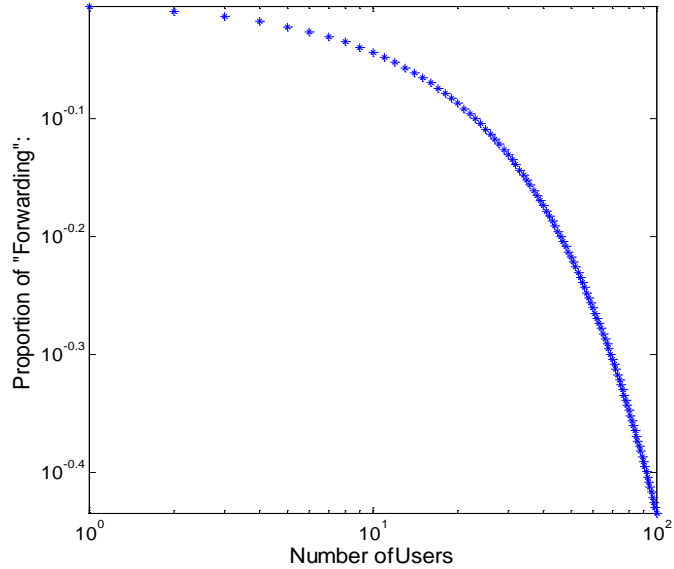

**Figure 1: An example of "Forwarding" Behaviors**

Based on the statistical analysis above, we design a probability distribution formula as below:

$$f(y_{ij}, Z_i, X_j(U)) = \alpha \times \frac{1}{Z_f} \times e^{-y_{ij} \times (Z_i(U) - X_j(U))^2} \quad (2)$$

$$g(y_{ij}, Z_i, X_j(TP)) = \beta \times \frac{1}{Z_g} \times e^{-y_{ij} \times (Z_i(TP) - X_j(TP))^2} \quad (3)$$

$$h(y_{ij}, Z_i, X_j(KW)) = \gamma \times \frac{1}{Z_h} \times e^{-y_{ij} \times (Z_i(KW) - X_j(KW))^2} \quad (4)$$

where $Z$ can be defined as the integration of meta-item, the definition of $Z$ is defined as below:

$$Z_f = \iint \alpha \times e^{-y_{ij} \times (Z_i(U) - X_j(U))^2} dZdX \quad (5)$$

$$Z_g = \iint \beta \times e^{-y_{ij} \times (Z_i(TP) - X_j(TP))^2} dZdX \quad (6)$$

$$Z_h = \iint \gamma \times e^{-y_{ij} \times (Z_i(KW) - X_j(KW))^2} dZdX \quad (7)$$

In order to incorporate Social Influence into the Factor Graph model, we design a indirect influence mechanism to build the social influence relationship among users, its main idea is based on Social Balance theory, which illustrates a transitive feature of social network. For two recommendation behaviors $W_i$ and $W_j$, we will build a connection between them if they have common topic $TP$. Then the indirect influence

between $W_i$ and $W_j$ can be described as below:

$$Graph(W_i, W_j) = \lambda \times \frac{1}{Z_G} \times e^{-(W_i(Y)-W_j(Y))^2} \quad (8)$$

Then the log-likehood of the built network $G$ can be described as below:

$$\log(P(Y^U | Y^K, G)) = \sum_{Y^U|Y^K,j} \log\{Graph(W_i,W_j)\} \sum_{Y^U|Y^K,i} \{\log[f(Y_i, W_i(Z), W_i(X(U)))] + \log[g(Y_i, W_i(TP), W_i(X(TP)))] + \log[h(Y_i, W_i(KW), W_i(X(KW)))]\} \quad (9)$$

In order to obtain the optimized value of the model, which could maximize the log-likelihood of formula (9), we design the vector $\phi = \{\alpha, \beta, \gamma, \lambda\}$, $S = \{Graph, f, g, h\}$ and $Z = \{Z_G, Z_\alpha, Z_\beta, Z_\gamma\}$. Let $\Omega = \log(P(Y^U | Y^K, G))$, then:

$$\Omega = \log \sum_{Y^U|Y^U \&\& Y^K} \frac{1}{Z} \{\phi^T S\} = \log \sum_{Y^U|Y^U \&\& Y^K} \{\phi^T S\} - \log Z$$
$$= \log \sum_{Y^U|Y^U \&\& Y^K} \{\phi^T S\} - \log \sum_{Y^U \&\& Y^K} \{\phi^T S\} \quad (10)$$

$$\frac{\partial \Omega}{\partial \phi} = \frac{\sum_{Y^U|Y^U \&\& Y^K} \{\phi^T S\} S}{\sum_{Y^U|Y^U \&\& Y^K} \{\phi^T S\}} - \frac{\sum_{Y^U \&\& Y^K} \{\phi^T S\} S}{\sum_{Y^U \&\& Y^K} \{\phi^T S\}} \quad (11)$$

In formula (11), we first need to prove that there exists optimization value, which can guarantee the convergence of the designed algorithm. First, for one extremely situation, which there is no unknown "Recommendation Behaviors" exists, which means that $|Y^U|=0$, then $\partial\Omega/\partial\phi = 0$. If there exists $|Y^U|\neq 0$, then if all unknown $Y^U$ can be correctly estimated, we can also obtain $\partial\Omega/\partial\phi \sim 0$. So the main purpose of the designed algorithm is how to obtain that optimized status and how to guarantee the efficiency of the proposed model (The proportion between $Y^U$ and $Y^K$). Thus the learned process can be adopted by applying Newton Gradient Descending Algorithm [18].

## 5. EXPERIMENT RESULTS

The proposed model for weibo recommendation is very general and can be applied to analyze different social networks. In this section, we present various experiments to evaluate the performance of the proposed approach.

**5.1 Experiment Setup**

**Data Sets:** We perform our experiment on Tencent QQ micro-blogging. The whole data set is collected from 2011.Nov.01 to 2012.Jan.05, which contains about 40,000,000 micro-blogs each day. To better evaluate out methods, we first all users according to their activities, the most active users with high number of "Forwarding" behaviors are chosen as experimental objects. Finally, we select 1,100 users from top 2,000 ranked most active users. The total statistical information can be summarized as below:

**Table 1: Experimental Data Summarization**

| User Number | Known behaviors | Unknown behaviors | Total Relationships | Key words |
|---|---|---|---|---|
| 1,100 | 292,316 | 165,053 | 1,551,621 | 110,000 |

**Comparison Method:** we compare the following methods for our proposed algorithm:

In our research, we use three classical algorithm: SVM (Support Vector Machine), Logistical Regression, CRF(Conditional Random Field) to make comparison. The main idea is to predict users' interest toward a certain weibo based on their historical behavior records. For SVM, we adopt SVM-light[2]; for Logistical Regression, we adopt Statistical tool box[3]; for CRF, we adopt the code provided by Tang, Zhuang [3] in 2011.

The basic learning algorithm is implemented using C++ and all experiments are performed on a servers cluster with 36 machines, each of which contains 15 Intel(R) Xeon(R) processors (2.13GHZ) and 60G memory. The data set is stored in HDFS system.

**5.2 Prediction Performance**

On all the training data sets, we use the historic users' behaviors to train the action tracking model and use the learned model to predict the users' opinions for different objects. The comparison results can be seen in Table 2:

**Table 2: Performance of Opinion prediction with different approaches**

|  | Accuracy | Recall | F1-Score |
|---|---|---|---|
| **TSI_WR** | 69.02% | 96.05% | 0.8088 |
| **SVM** | 31.88% | 100% | 0.4835 |
| **LR** | 66.0896% | 100% | 0.7958 |
| **CRF** | 68.99% | 96.12% | 0.8088 |

As can be seen in Table 2, our proposed model TSI_WR gains the highest F1-Score than SVM and CRF, it has a higher accuracy score than CRF, while its recall is a little lower than CRF, which means that it can train a more accurate rules to judge the uncertain situation, for example, TSI_WR will drop those nodes with high uncertainty, while CRF will keep them, which leads to a higher probability of error occurring. Another reason is that by considering "Indirect Influence", we could make recommendation for those users without direct connection by applying TSI_WR model, while for CRF, it will occur mistakes for those situations.

In another aspect, we would like to consider for all of those features, which one can gain a big contribution to the performance of our proposed model, we design the experiment as follow:
- For each time, we omit one attribute from original TSI_WR model and run it on the Training and Testing data;
- We calculate Accuracy, Recall and F1-Score for each trained results and make comparison analysis for all results.

The experiment results can be seen in Table 3 as below:

---
[2] http://svmlight.joachims.org/
[3] http://luna.cas.usf.edu/

Table 3: The Contribution of each attribute of TSI_WR

| Item | ALL | No Edge | No FN | No KW | No TP | No RN | No CN |
|---|---|---|---|---|---|---|---|
| **Accuracy** | 0.6902 | 0.6899 | **0.6702** | 0.6820 | 0.6822 | 0.6746 | 0.6903 |
| **Precision** | 0.6985 | 0.6981 | 0.7026 | 0.6907 | **0.6822** | 0.7086 | 0.6986 |
| **Recall** | 0.9605 | 0.9612 | 0.8958 | 0.9669 | 1.0000 | **0.8884** | 0.9605 |
| **F1-Score** | 0.8088 | 0.8088 | **0.7875** | 0.8058 | 0.8111 | 0.7883 | 0.8089 |

As can be seen in Table 3, the contribution of *FN* is the biggest than others, which implies that it is a relatively important index to have latent correlation with users' interesting preference; *RN* is also an effective factor to reflect the latent relationship between users. "Edge" which represents "Indirect Influence" also contributes significant improvement in the experiment, which means that the assumption of the existing of "indirect influence" is established, but due to the limitation of data sparsity, the improvement is not as much as our expectations.

## 6. CONCLUSION

In this paper, we propose a TSI_WR model for solving online weibo recommendation problem in Tencent, the largest micro-blogging services in China. Different with many previous studies, our algorithm first consider applying semi-supervised algorithm to obtain a high performance under the condition of data sparsity; second consider to incorporate "Social Influence" into a multi-attributes Factor Graph model, which could detect the indirect influence among Tencent users and improve the efficiency of model prediction; third, we use TSI_WR model to analyze the contribution of different features, and calculate the importance of each one, which could provide a good foundation for feature selection in the future.